\documentclass[english]{article}
\usepackage[T1]{fontenc}
\usepackage[latin9]{inputenc}
\usepackage{geometry}
\usepackage{textcomp}
\usepackage{amstext}
\usepackage{amssymb}
\usepackage{esint}

\makeatletter

\date{}

\makeatother

\usepackage{babel}
\begin{document}

\title{CP(N) model on regions with boundary}

\maketitle
\begin{center} { A. Pikalov\footnote{arseniy.pikalov@phystech.edu}} \\ 
\emph{Institute of Theoretical and Experimental Physics,}\\ 
\emph{Moscow 117218, Russia}  \\ 
\emph{Moscow Institute of Physics and Technology,} \\ 
\emph{Dolgoprudny 141700, Russia } 
\end{center} 
\begin{abstract}
In this note we discuss the CP(N) model in
large $N$ limit in saddle point approximation on disc and annulus
with various combinations of Dirichlet and Neumann boundary conditions.
We show that homogeneous condensate is not a saddle point in any of
considered cases. Behavior of inhomogeneous condensate near boundary
is briefly discussed.
\end{abstract}

\section{Introduction}

Two-dimensional $\mathbb{C}P\left(N\right)$ model in the large $N$
limit was solved in \cite{adda,witten} (see \cite{sigma model} for
a detailed review). It was found that the model is asymptotically
free and exhibits dynamical generation of mass via dimensional transmutation,
which makes it similar to QCD. The $\mathbb{C}P\left(N\right)$ model
appears as an effective low energy theory for worldsheet of non-Abelian
string. It was found in supersymmetric case in \cite{hanany 1,konishi,shifman yung,hanany 2}
and then in non-supersymmetric case in \cite{gorsky 1}. Both supersymmetric
and non-supersymmetric versions of the model on interval with periodical
boundary conditions were studied in \cite{monin shifman}. Recently
this theory on finite interval and on disc was discussed \cite{milekhin 1,milekhin 2,bolognesi 1,bolognesi 2,nitta 2,gorsky 2}.
It was shown that for a finite interval with Dirichlet boundary conditions
a homogeneous condensate is not a solution of saddle point equation
\cite{bolognesi 1,bolognesi 2}. In \cite{milekhin 2} a modification
of boundary conditions consistent with homogeneous solution was proposed.
Namely, instead of $\mathbb{C}P\left(N\right)$ model $\mathbb{C}P\left(2N\right)$
model with $N$ components satisfying Dirichlet boundary conditions
and $N$ components satisfying Neumann boundary conditions was suggested.
This choice of boundary conditions leads to full cancellation of coordinate
dependance in saddle point equation and thus to possibility of homogeneous
solution. Similar method was successfully applied to Grassmannian
model on the finite interval \cite{grassmannian}. A short time ago
an exact solution based on correspondence with Gross-Neveu model was
proposed \cite{nitta 2}. Details of the correspondence are presented
in \cite{nitta 1}. Behavior of this solution near boundary is different
from one found in \cite{bolognesi 1,bolognesi 2}. The model on the
disc was discussed in \cite{gorsky 2} in connection with problem
of non-Abelian string decay. However, the case of inhomogeneous condensate
was not analyzed. In this note we will focus on inhomogeneous condensate
on disc and annulus. Our main result is that there is no choice of
boundary conditions compatible with homogeneous condensate. We check
this conclusion for some types of boundary conditions and expect that
similar analysis can be applied to more complicated choice of boundary
conditions. Nevertheless, we can find boundary conditions similar
to one used in \cite{milekhin 2} for which the leading terms in the
divergences of condensate near boundary cancel.

This note is organized as follows. In the Section 2 we revise the
one-loop effective action and the gap equation for $\mathbb{C}P\left(N\right)$
model. In the Section 3 we discuss condensate on disc. We firstly
consider Dirichlet boundary conditions and then discuss combination
of Dirichlet and Neumann boundary conditions similar to one used in
\cite{milekhin 2}. In the Section 4 similar analysis is performed
for the case of annulus.

\section{Effective action and gap equation}

The Euclidean action for $\mathbb{C}P\left(N\right)$ $\sigma$ model
is
\begin{equation}
S=\int d^{2}x\left(\left(D_{\mu}n_{i}\right)^{*}\left(D_{\mu}n_{i}\right)+\lambda\left(\left|n\right|^{2}-r\right)\right),\;i=1,\dots,\,N+1
\end{equation}
We presume that number of components $N$ is large. Here $\lambda$
is Lagrange multiplier that leads to constraint $n_{i}n^{i}=r$. For
the first component the boundary condition is chosen to be consistent
with constraint $n_{1}=\sqrt{r}$. On fields $n_{i}$ with $i=2,\dots,\,N+1$
we impose Dirichlet or Neumann boundary conditions which will be specified
later. $D_{\mu}=\partial_{\mu}-iA_{\mu}$ is covariant derivative.
We do not consider dynamics of gauge field $A_{\mu}$ and suppose
that its vacuum expectation value is zero in the leading order of
$1/N$ expansion. At classical level it is just a dummy field that
can be eliminated by its equation of motion. In the later analysis
it is assumed that $A_{\mu}=0$ and $n_{1}=\sigma$ is real. We consider
the theory on the disc ($\sqrt{x_{0}^{2}+x_{1}^{2}}<R$) and annulus
$R_{1}<\sqrt{x_{0}^{2}+x_{1}^{2}}<R_{2}$ with different boundary
conditions. As usual, we can integrate over components with trivial
boundary conditions $n_{i},\;i=2,\dots,\,N+1$ and obtain effective
action in terms of fields $\lambda$ and $\sigma$. Thus effective
action
\begin{equation}
S_{eff}=NTr\log\left(-\partial^{2}+\lambda\right)+\int d^{2}x\left(\left(\partial\sigma\right)^{2}+\lambda\left(\sigma^{2}-r\right)\right)
\end{equation}
 Variation of the action with respect to $\lambda$ yields the gap
equation 
\begin{equation}
NTr\frac{\delta\lambda}{-\partial^{2}+\lambda}+\int d^{2}x\delta\lambda\left(\sigma^{2}-r\right)
\end{equation}
We can express this equation through the eigenfunction of operator
$-\partial^{2}+\lambda$:

\begin{equation}
\left(-\partial^{2}+\lambda\right)f_{\alpha}\left(x\right)=\kappa_{\alpha}f_{\alpha}
\end{equation}

\begin{equation}
N\sum_{\alpha}\frac{\left|f_{\alpha}\left(x\right)\right|^{2}}{\kappa_{\alpha}}+\sigma^{2}-r=0,\label{eq:gap1}
\end{equation}
Here $f_{\alpha}$must satisfy the same boundary conditions as corresponding
field $n_{i}$ and normalization condition $\int d^{2}x\left|f_{\alpha}\left(x\right)\right|^{2}=1$,
integration is over the considered region. Throughout this section
$\alpha$ is index that enumerates all eigenvalues of particular differential
operator so summation in (\ref{eq:gap1}) is over all eigenfunctions.
Note that the sum is Green function of the operator $-\partial^{2}+\lambda$
taken at coinciding points.
\begin{equation}
\left(-\partial_{x}^{2}+\lambda\right)G\left(x,\,y\right)=\delta\left(x-y\right),\;G\left(x,\,y\right)=\sum_{\alpha}\frac{f_{\alpha}\left(x\right)f_{\alpha}^{*}\left(y\right)}{\kappa_{\alpha}}\label{eq:greenfn}
\end{equation}
so 
\begin{equation}
\sum_{\alpha}\frac{\left|f_{\alpha}\left(x\right)\right|^{2}}{\kappa_{\alpha}}=\lim_{y\to x}G\left(x,\,y\right)
\end{equation}
Strictly speaking, this limit is infinite, so we need to regularize
the sum by considering small but finite distance between points $x$
and $y$. The equation obtained by variation of action with respect
to $\sigma$ is 
\begin{equation}
\left(-\partial^{2}+\lambda\right)\sigma=0\label{eq:sigma}
\end{equation}
The problem has rotational symmetry, so we assume that all fields
depend only on the distance to the center:
\[
\lambda=\lambda\left(\rho\right),\;\sigma=\sigma\left(\rho\right),\;f=\exp\left(il\varphi\right)g\left(\rho\right),
\]
\[
\text{where }x_{1}=\rho\cos\varphi,\;x_{2}=\rho\sin\varphi.
\]
The sum over eigenfunctions in the equation (\ref{eq:gap1}) is divergent.
However we can subtract from eigenfunctions their mean values and
obtain conditionally convergent sum which contains all information
about dependence on the coordinates and divergent sum which leads
to renormalization of $r$. The transformed equation is

\begin{equation}
N\sum\frac{1}{\kappa_{\alpha}}\left(\left|f_{\alpha}\left(x\right)\right|^{2}-\frac{1}{A}\right)+\sigma^{2}+\frac{N}{A}\sum_{\alpha}\frac{1}{\kappa_{\alpha}}-r=0\label{eq:gap2}
\end{equation}
Here $A$ is area of the region, $A=\pi R^{2}$ for a disk and $A=\pi\left(R_{2}^{2}-R_{1}^{2}\right)$
for annulus. Now the first sum converges. It will be used in next
sections to investigate the behavior of $\sigma$ near boundary. We
can not solve the system of equation (\ref{eq:gap1}) and (\ref{eq:sigma})
analytically. However it is possible to calculate eigenvalues and
eigenfunctions for Dirichlet and Neumann boundary conditions for the
case $\lambda\approx const=m^{2}$ in terms of zeros of Bessel functions.
We assume that large eigenvalues for arbitrary $\lambda$ are almost
the same, so this allows to understand behavior of $\sigma$ near
boundaries using this eigenvalues and eigenfunctions. So we are interested
in the behavior of the sum
\begin{equation}
\Sigma\left(x\right)=\sum_{\alpha}\left(\frac{1}{A}-\frac{\left|f_{\alpha}\left(x\right)\right|^{2}}{\kappa_{\alpha}}\right)\label{eq:sum}
\end{equation}
near the boundary, where $f_{\alpha}$ are eigenfunctions of operator
$-\partial^{2}+m^{2}$. Our hypothezis is that near the boundary $\sigma^{2}\sim N\Sigma$
at least in case when this sum tends to infinity as we approach to
boundary so by considering (\ref{eq:sum}) we can find out whether
$\sigma$ is finite near the boundary or not.

\section{Model on disc}

Firstly we consider model on disc with Dirichlet boundary conditions.
The eigenfunctions and eigenvalues are 
\begin{equation}
f_{l,\,n}=A_{l,\,n}\exp\left(il\varphi\right)J_{l}\left(\mu_{l,\,n}\frac{\rho}{R}\right),\;\kappa_{l,\,n}=\frac{\mu_{l,\,n}^{2}}{R^{2}}+m^{2}
\end{equation}
Here $J_{l}\left(z\right)$ is Bessel function of the first kind,
which is regular at $z=0$ and $J_{l}\left(\mu_{l,\,n}\right)=0.$
Normalization constant can be found from condition
\begin{equation}
A_{l,\,n}^{2}\pi R^{2}J_{l-1}^{2}\left(\mu_{l,\,n}\right)=1
\end{equation}
so the sum (\ref{eq:sum}) becomes 
\begin{equation}
\Sigma_{D}\left(\rho\right)=\frac{1}{\pi}\sum_{l=-\infty}^{\infty}\sum_{n=1}^{\infty}\frac{1}{\mu_{l,\,n}^{2}+m^{2}R^{2}}\left(1-\frac{J_{l}^{2}\left(\mu_{l,\,n}\frac{\rho}{R}\right)}{J_{l-1}^{2}\left(\mu_{l,\,n}\right)}\right)\label{eq:sumd}
\end{equation}
In the same way for Neumann boundary conditions we obtain
\begin{equation}
f_{l,\,n}=B_{l,\,n}\exp\left(il\varphi\right)J_{l}\left(\tilde{\mu}_{l,\,n}\frac{\rho}{R}\right),\;\kappa_{l,\,n}=\frac{\tilde{\mu}_{l,\,n}^{2}}{R^{2}}+m^{2}
\end{equation}
where $J_{l}^{\prime}\left(\tilde{\mu}_{l,\,n}\right)=0$ and 
\begin{equation}
B_{l,\,n}^{2}\pi R^{2}\left(J_{l}^{2}\left(\tilde{\mu}_{l,\,n}\right)-J_{l-1}^{2}\left(\tilde{\mu}_{l,\,n}\right)\right)=1
\end{equation}
so the sum is 
\begin{equation}
\Sigma_{N}\left(\rho\right)=\frac{1}{\pi}\sum_{l=-\infty}^{\infty}\sum_{n=1}^{\infty}\frac{1}{\mu_{l,\,n}^{2}+m^{2}R^{2}}\left(1-\frac{J_{l}^{2}\left(\tilde{\mu}_{l,\,n}\frac{\rho}{R}\right)}{J_{l}^{2}\left(\tilde{\mu}_{l,\,n}\right)-J_{l-1}^{2}\left(\tilde{\mu}_{l,\,n}\right)}\right)\label{eq:sumn}
\end{equation}
Both sums have logarithmic divergence near the boundary. The easiest
way to see this is consideration of Green function (\ref{eq:greenfn}).
Numerical computation of this sums confirms this conclusion. The leading
terms obtained from Green functions are 
\begin{equation}
\Sigma_{D}\left(\rho\right)=-\Sigma_{N}\left(\rho\right)=\frac{1}{2\pi}\log\left(\frac{1}{R-\rho}\right)
\end{equation}

We can hope to cancel this divergence by considering $\mathbb{C}P\left(2N\right)$
model and imposing on $N$ fields Dirichlet boundary conditions and
on other $N$ fields Neumann boundary conditions. Thus instead of
sums (\ref{eq:sumd}) and (\ref{eq:sumn}) we will have 
\begin{equation}
\Sigma\left(x\right)=\Sigma_{D}\left(\rho\right)+\Sigma_{N}\left(\rho\right)
\end{equation}
This sum does not have logarithmic divergence. However, asymptotic
of zeros of Bessel function and its derivative are (the first of this
expressions is a case of general formula 8.547 from \cite{table},
the second can be derived similarly)
\begin{equation}
\mu_{l,\,n}=\left(n+\frac{l}{2}-\frac{1}{4}\right)\pi-\frac{4l^{2}-1}{8\pi\left(n+l/2-1/4\right)}
\end{equation}
\begin{equation}
\tilde{\mu}_{l,\,n}=\left(n+\frac{l}{2}+\frac{1}{4}\right)\pi-\frac{4l^{2}+3}{8\pi\left(n+l/2+1/4\right)}
\end{equation}
so $\tilde{\mu}_{l,\,n}-\mu_{l,\,n}\to\pi/2$ as $n\to\infty$ and
frequencies of oscillations of eigenfunctions with Dirichlet and Neumann
boundary conditions are different, so we might expect slow oscillations
of amplitude of the sum so $\Sigma\ne const$. This conclusions are
also confirmed by numerical summation. 

\section{Model on annulus}

Now let us consider $\mathbb{C}P\left(2N\right)$ model on annulus.
We want to impose different types of boundary conditions on $n_{i}$
for $i=2,\dots,\,N+1$ and for $i=N+2,\dots,\,2N+1$ in order to cancel
divergence of $\Sigma$ at the boundary. One of the possible choices
is to use Dirichlet boundary condition for $N$ fields and Neumann
boundary conditions for other $N$ fields. In both of this cases eigenfunctions
are
\begin{equation}
f_{l,\,n}=A_{l,\,n}\exp\left(il\varphi\right)\left(J_{l}\left(\mu_{l,\,n}\frac{\rho}{R_{2}}\right)\cos\alpha_{l,\,n}-Y_{l}\left(\mu_{l,\,n}\frac{\rho}{R_{2}}\right)\sin\alpha_{l,\,n}\right)
\end{equation}
where $Y_{l}\left(z\right)$ is Bessel function of the second kind,
$\alpha_{l,\,n}\in\left[0,\,\pi\right]$and $\mu_{l,\,n}$ and the
parameter $\alpha_{l,\,n}$ are determined from boundary conditions.
Corresponding eigenvalue is
\begin{equation}
\kappa_{l,\,n}=\frac{\mu_{l,\,n}^{2}}{R_{2}^{2}}+m^{2}
\end{equation}
We will use following formulas for large zeros of function $J_{l}\left(z\right)\cos\alpha_{l,\,n}-Y_{l}\left(z\right)\sin\alpha_{l,\,n}$
(this expression is formula 8.547 from \cite{table} again)
\begin{equation}
z_{l,\,n}=\left(n+\frac{l}{2}-\frac{1}{4}\right)\pi-\alpha_{l,\,n}-\frac{4l^{2}-1}{8\left\{ \pi\left(n+l/2-1/4\right)-\alpha_{l,\,n}\right\} }
\end{equation}
and zeros of its derivative
\begin{equation}
\tilde{z}_{l,\,n}=\left(n+\frac{l}{2}-\frac{1}{4}\right)\pi-\alpha_{l,\,n}-\frac{4l^{2}+3}{8\left\{ \pi\left(n+l/2-1/4\right)-\alpha_{l,\,n}\right\} }
\end{equation}
We introduce dimensionless parameter $\gamma=R_{1}/R_{2}$. Firstly
consider Dirichlet boundary conditions
\[
n\left(R_{1}\right)=n\left(R_{2}\right)=0
\]
We are going to find asymptotic of eigenvalues with large $n.$ From
boundary conditions follow equation for $\alpha$
\begin{equation}
\mu_{l,\,n}\gamma=\left(n_{1}+\frac{l}{2}-\frac{1}{4}\right)\pi-\alpha_{l,\,n}-\frac{4l^{2}-1}{8\left\{ \pi\left(n_{1}+l/2-1/4\right)-\alpha_{l,\,n}\right\} }
\end{equation}
\begin{equation}
\mu_{l,\,n}=\left(n_{2}+\frac{l}{2}-\frac{1}{4}\right)\pi-\alpha_{l,\,n}-\frac{4l^{2}-1}{8\left\{ \pi\left(n_{2}+l/2-1/4\right)-\alpha_{l,\,n}\right\} }
\end{equation}
Here $n_{2}-n_{1}=n$ because in one dimension the number of the root
is equal to number of zeros of eigenfunction minus one due to oscillatory
theorem. So for the large $n$ we have approximately 
\begin{equation}
\mu_{l,\,n}=\frac{\pi n}{1-\gamma}
\end{equation}
\begin{equation}
n_{1}=\frac{\gamma n}{1-\gamma}-\left(\frac{l}{2}-\frac{1}{4}-\frac{\alpha_{l,\,n}}{\pi}\right),\;n_{2}=\frac{\gamma n}{1-\gamma}-\left(\frac{l}{2}-\frac{1}{4}-\frac{\alpha_{l,\,n}}{\pi}\right)
\end{equation}
$\alpha_{l,\,n}$ can be determined from the condition that $n_{1}$is
integer. Thus we can calculate correction for the $\mu_{l,\,n}$
\begin{equation}
\mu_{l,\,n}=\frac{\pi n}{1-\gamma}+\frac{4l^{2}-1}{8\pi n}\frac{1+\gamma}{\gamma}
\end{equation}
So approximate eigenvalues for large $n$ are 
\begin{equation}
\kappa_{l,\,n}=\left(\frac{\pi n}{R_{2}-R_{1}}+\frac{4l^{2}-1}{8\pi n}\left(\frac{1}{R_{1}}+\frac{1}{R_{2}}\right)\right)^{2}+m^{2}\label{eq:ev1}
\end{equation}
Similarly for the Neumann boundary conditions 
\[
\partial_{\rho}n\left(R_{1}\right)=\partial_{\rho}n\left(R_{2}\right)=0
\]
eigenvalues are

\begin{equation}
\tilde{\mu}_{l,\,n}=\frac{\pi n}{1-\gamma}+\frac{4l^{2}+3}{8\pi n}\frac{1+\gamma}{\gamma}
\end{equation}

\begin{equation}
\tilde{\kappa}_{l,\,n}=\left(\frac{\pi n}{R_{2}-R_{1}}+\frac{4l^{2}+3}{8\pi n}\left(\frac{1}{R_{1}}+\frac{1}{R_{2}}\right)\right)^{2}+m^{2}\label{eq:ev2}
\end{equation}
The first term in brackets in (\ref{eq:ev1}) and (\ref{eq:ev2})
corresponds to the case of an interval. For large $l$ eigenvalues
differ significantly from the case of interval. However, one can easily
obtain that
\[
\tilde{\mu}_{l,\,n}-\mu_{l,\,n}=\frac{1}{4\pi n}\frac{1+\gamma}{\gamma}\to0,\;\text{as }n\to\infty
\]
Note that the leading term of asymptotic does not contain the large
number $l$. It means that differences of spacial frequencies of eigenfunctions
tend to zero uniformly so we might expect better cancellation of divergence
than for the case of disc. 

\section{Discussion}

In this note we have discussed $\mathbb{C}P\left(N\right)$ on disc
and annulus in large $N$ limit. It was shown that on disc with Dirichlet
boundary conditions homogeneous solution is impossible. We also showed
that mixing of Dirichlet and Neumann conditions does not yield constant
in $\sigma$ condensate neither in case of disc nor in case of annulus.
However we claim that this combination of boundary condition makes
$\sigma$ finite near boundary. It might be useful for numerical analysis
of the inhomogeneous condensates.

\section*{Acknowledgment}

We are grateful to A. Gorsky for suggesting this problem and numerous
discussions.


\begin{thebibliography}{10}
\bibitem{adda}A. D\textquoteright Adda, M. Luscher, and P. Di Vecchia.
\textquotedblleft A 1/N Expandable Series of Nonlinear Sigma Models
with Instantons\textquotedblright . In: Nucl. Phys. B146 (1978), pp.
63\textendash 76. DOI: 10.1016/0550-3213(78)90432-7.

\bibitem{witten}Edward Witten. \textquotedblleft Instantons, the
quark model, and the 1/N expansion\textquotedblright . In: Nucl. Phys.
B 149.2 (1979), pp. 285 \textendash 320. DOI: 10.1016/0550-3213(79)90243-8.

\bibitem{sigma model}V.A. Novikov, Mikhail A. Shifman, A.I. Vainshtein,
Valentin I. Zakharov. ``Two-Dimensional Sigma Models: Modeling Nonperturbative
Effects of Quantum Chromodynamics''. In: Phys.Rept. 116 (1984) 103.
DOI: 10.1016/0370-1573(84)90021-8.

\bibitem{milekhin 1}A. Milekhin. ``CP(N-1) model on finite interval
in the large N limit''. In: Phys.Rev. D86 (2012) 105002. DOI: 10.1103/PhysRevD.86.105002.
arXiv:1207.0417 {[}hep-th{]}. 

\bibitem{milekhin 2}A. Milekhin. ``CP(N) sigma model on a finite
interval revisited''. In: Phys.Rev. D95 (2017) no.8, p. 085021. DOI:
10.1103/PhysRevD.95.085021. arXiv:1612.02075 {[}hep-th{]}.

\bibitem{grassmannian}Dmitriy Pavshinkin. ``Grassmannian sigma model
on a finite interval''. arXiv:1708.06399 {[}hep-th{]}. 

\bibitem{bolognesi 1}Stefano Bolognesi, Kenichi Konishi, Keisuke
Ohashi. \textquotedblleft Large-N CP(N\textminus 1) sigma model on
a finite interval\textquotedblright . In: JHEP 10 (2016), p. 073.
DOI: 10.1007/JHEP10(2016) 073. arXiv: 1604.05630 {[}hep-th{]}.

\bibitem{bolognesi 2}Alessandro Betti, Stefano Bolognesi, Sven Bjarke
Gudnason, Kenichi Konishi, Keisuke Ohashi. ``Large-N CP(N-1) sigma
model on a finite interval and the renormalized string energy''.
arXiv:1708.08805 {[}hep-th{]}

\bibitem{nitta 1}Muneto Nitta, Ryosuke Yoshi. ``Self-Consistent
Exact Solutions of Inhomogeneous Condensates in Quantum CP(N\textminus 1)
Model''. arXiv:1707.03207 {[}hep-th{]}.

\bibitem{nitta 2}Antonino Flachi, Muneto Nitta, Satoshi Takada, Ryosuke
Yoshii. ``Casimir Force for the CP(N\textminus 1) Model''. arXiv:1708.08807
{[}hep-th{]}. 

\bibitem{hanany 1}Amihay Hanany, David Tong. ``Vortices, instantons
and branes''. In: JHEP 0307 (2003) 037. DOI: 10.1088/1126-6708/2003/07/037.
{[}arXiv:hep-th/0306150{]}. 

\bibitem{konishi}Roberto Auzzi, Stefano Bolognesi, Jarah Evslin,
Kenichi Konishi, Alexei Yung. ``NonAbelian superconductors: Vortices
and confinement in N=2 SQCD''. In: Nucl.Phys. B673 (2003) 187-216.
DOI: 10.1016/j.nuclphysb.2003.09.029. {[}arXiv:hep-th/0307287{]}.

\bibitem{shifman yung}M. Shifman, A. Yung. ``NonAbelian string junctions
as confined monopoles''. In: Phys.Rev. D70 (2004) 045004. DOI: 10.1103/PhysRevD.70.045004.
{[}arXiv:hep-th/0403149{]}.

\bibitem{hanany 2}Amihay Hanany, David Tong. ``Vortex strings and
four-dimensional gauge dynamics''. In: JHEP 0404 (2004) 066. DOI:
10.1088/1126-6708/2004/04/066. {[}arXiv:hep-th/0403158{]} 

\bibitem{gorsky 1}A. Gorsky, M. Shifman, A. Yung. ``Non-Abelian
Meissner effect in Yang-Mills theories at weak coupling ''. In: Phys.Rev.
D71 (2005) 045010. DOI: 10.1103/PhysRevD.71.045010. {[}arXiv:hep-th/0412082{]}. 

\bibitem{gorsky 2}A. Gorsky, A. Milekhin. ``The CP(N \textminus{}
1) model on a Disc and Decay of a Non-Abelian String''. In: Phys.Rev.
D88 (2013) no.8, 085017. DOI: 10.1103/PhysRevD.88.085017. arXiv:1306.3565
{[}hep-th{]}.

\bibitem{monin shifman}Sergey Monin, Mikhail Shifman, Alexei Yung.
``Non-Abelian String of a Finite Length''. In: Phys.Rev. D92 (2015)
no.2, 025011. DOI: 10.1103/PhysRevD.92.025011. arXiv:1505.07797 {[}hep-th{]}.

\bibitem{table}I.S. Gradshteyn, I.M. Ryzhik. Table of Integrals,
Series, and Products.
\end{thebibliography}
\end{document}